\documentclass{PoS}
\usepackage{lineno}
\title{Using muon rings for the optical throughput calibration of the SST-1M prototype for the Cherenkov Telescope Array}

\ShortTitle{Calibration of the SST-1M using muon rings}

\author{\speaker{S.~Toscano} $^{a, n}$, E. Prandini$^{a}$\\
        E-mail: \email{simona.toscano@vub.ac.be} \\
W.~Bilnik$^{k}$,
J. B\l{}ocki$^{c}$,
L.~.Bogacz$^{m}$,
T~.Bulik$^{d}$,
F.~Cadoux$^{b}$,
A.~Christov$^{b}$,
M.~Cury{\l}o$^{c}$,
D.~della Volpe$^{b}$,
M.~Dyrda$^{c}$,
Y.~Favre$^{b}$,
A.~Frankowski$^{g}$,
\L{}. Grudniki$^{c}$,
M.~Grudzi{\'n}ska$^{d}$,
M. Heller$^{b}$,
B.~Id{\'z}kowski$^{e}$,
M.~Jamrozy$^{e}$,
M.~Janiak$^{g}$,
J.~Kasperek$^{k}$,
K.~Lalik$^{k}$,
E. Lyard$^{a}$,
E.~Mach$^{c}$,
D.~Mandat$^{l}$,
A.~Marsza{\l}ek$^{c,e}$,
J.~Micha{\l}owski$^{c}$,
R.~Moderski$^{g}$,
T.~Montaruli$^{b}$,
A. Neronov$^{a}$,
J.~Niemiec$^{c}$,
M.~Ostrowski$^{e}$,
P.~Pa{\'s}ko$^{f}$,
M.~Pech$^{l}$,
A.~Porcelli$^{b}$,
P.~Rajda$^{k}$,
M.~Rameez$^{b}$,
E. Jr. Schioppa$^{b}$,
P.~Schovanek$^{l}$,
K.~Seweryn$^{f}$, 
K.~Skowron$^{c}$,
V.~Sliusar$^{j}$,
M.~Sowi{\'n}ski$^{c}$,
{\L}.~Stawarz$^{e}$,
M.~Stodulska$^{e}$,
M.~Stodulski$^{c}$,
I.~Troyano Pujadas$^{b}$, 
R.~Walter$^{a}$,
M.~Wi{\c e}cek$^{k}$,
A.~Zagda\'{n}ski$^{e}$,
K.~Zi{\c e}tara$^{e}$,
P.~{\.Z}ychowski$^{c}$ for the CTA Consortium\footnote{Full consortium author list at http://cta-observatory.org} \\
\footnotesize{
a. ISDC, Observatoire de Gen\`eve, Universit\'e de Gen\`eve, 1290 Versoix, Switzerland.\\
b. D\'epartment de physique nucleaire et corpusculaire, Universit\'e de Gen\`eve, CH-1205 Switzerland.\\
c. Instytut Fizyki J{\c a}drowej im. H. Niewodnicza{\'n}skiego Polskiej Akademii Nauk,  31-342 Krak{\'o}w, Poland.\\
d. Astronomical Observatory, University of Warsaw, Al. Ujazdowskie 4, 00-478 Warsaw, Poland\\
e. Astronomical Observatory, Jagiellonian University, ul. Orla 171, 30-244, Krak{\'o}w, Poland.\\
f. Centrum Bada{\'n} Kosmicznych Polskiej Akademii Nauk,  18a Bartycka str., 00-716 Warsaw, Poland.\\
g. Nicolaus Copernicus Astronomical Center, Polish Academy of Sciences,  Warsaw, Poland.\\
j. Astronomical Observatory, Taras Shevchenko Nat. University of Kyiv, Observatorna str., 3, Kyiv, Ukraine.\\
k. AGH University of Science and Technology, al.Mickiewicza 30, Krak{\'o}w, Poland,\\
l. Institute of Physics of the Czech Academy of Sciences, Prague, Czech Republic.\\
m. Department of Information Technologies, Jagiellonian University, 30-348 Krak{\'o}w, Poland.\\
n. Vrije Universiteit Brussels, Pleinlaan 2 1050 Brussels, Belgium.\\}
}

\abstract{Imaging Atmospheric Cherenkov Telescopes (IACTs) are ground-based instruments devoted to the study of very high energy gamma-rays coming from space. The detection technique consists of observing images created by the Cherenkov light emitted when gamma rays, or more generally cosmic rays, propagate through the atmosphere. While in the case of protons or gamma-rays the images present a filled and more or less elongated shape, energetic muons penetrating the atmosphere are visualised as characteristic circular rings or arcs. A relatively simple analysis of the ring images allows the reconstruction of all the relevant parameters of the detected muons, such as the energy, the impact parameter, and the incoming direction, with the final aim to use them to calibrate the total optical throughput of the given IACT telescope. We present the results of preliminary studies on the use of images created by muons as optical throughput calibrators of the single mirror small size telescope prototype SST-1M proposed for the Cherenkov Telescope Array.}

\FullConference{The 34th International Cosmic Ray Conference,\\
		30 July- 6 August, 2015\\
		The Hague, The Netherlands}

\begin{document}

\section{Introduction}

The Cherenkov Telescope Array project (CTA) \cite{CTA} will be the next generation ground based observatory in gamma-ray astronomy. It will consist of two arrays of imaging atmospheric Cherenkov telescopes (IACTs) installed in the two hemispheres to cover the full sky. In its final configuration, the Southern site of CTA will have more than a hundred telescopes of three different sizes arranged over an area of the order of few km$^2$. The array has been designed in such a way that it will be able to cover the wide energy range between few tens of GeV and at least 300 TeV, with an improved sensitivity by about a factor of 10 (at 1 TeV) compared to the existing experiments. The array of small-size telescopes (SSTs) will be dedicated to the observation of the most extreme gamma-ray sources emitting in energies between a few TeV and 300 TeV.

The SST-1M project is a Consortium of several Swiss and Polish institutes working on the design and construction of one of the prototypes for the SSTs of CTA \cite{SST}. The SST-1M prototype is the only SST telescope designed with one mirror. It uses the standard and  well-proven Davies-Cotton design for the optics and telescope structure and an innovative camera using Silicon-Photomultipliers (SiPM), inspired by the experience of the FACT camera \cite{FACT}. Thanks to their wide field-of-view (FoV) of about 9$^\circ$, the SSTs are particularly well suited for spectral studies of Galactic sources and Galactic Plane Surveys. The prototype of the SST-1M mechanical structure has been installed on a test site at the Institute of Nuclear Physics PAS in Krak\'ow in November 2013. The design of the camera has been finalized and the commissioning phase has already started \cite{CAMERA}.

The detection technique used by the IACTs consists in observing images created by the Cherenkov light emitted by electromagnetic showers induced in the atmosphere by gamma rays, or more generally cosmic rays. To analyse the observed gamma-ray characteristics, an accurate calibration of the camera \cite{CAMCALIB} as well as the evaluation of the instrument optical efficiency need to be performed. Highly energetic muons penetrating the atmosphere are visualised as characteristic circular rings or arcs in the camera plane and they represent a powerful and precise method to calibrate the optical throughput of IACTs \cite{CCF}.

In this paper we study the possibility of using the analysis of muon ring images as a calibrator for the optical throughput of the SST-1M prototype for the Cherenkov Telescope Array. 

\section{Muon images in IACTs}

Isolated muons produce sharply defined ring-like images in the focal plane and provide a powerful tool for monitoring the behaviour of the telescope performance characteristics like the point spread function (PSF) and the overall light collection efficiency. Muons passing through the centre of the telescope with trajectories parallel to its optical axis will produce azimuthally symmetric rings in the camera. The rings will have radii given by the Cherenkov angle of the muons and the total number of photons expected in the ring, mainly related to the muon energy, can be calculated from the measured value of this angle. The correspondence between the Cherenkov angle and the muon energy can be established following the Cherenkov equation:
\begin{equation}
\cos\theta_c = \frac{1}{n \cdot \sqrt{1 - (E_0/E_\mu)^2 }} \quad,
\label{eq:cherenkov}
\end{equation}
\noindent
where $E_0 \sim 0.105$~GeV is the muon rest mass, and $n$ the refractive index of air. The typical ring radius ($R$) is determined by the muon spectrum at the observatory altitude, e.g. $R\sim 1.2^\circ$ at 1500~m a.s.l. for $E_\mu = 20$~GeV. Muons arriving under a certain angle with respect to the telescope's axis will form rings or arcs with centres that are offset from the centre of the camera, while muons with non-zero impact parameters\footnote{The impact parameter is defined as the distance from the centre of the mirror to the impact point of the muon in the plane normal to the optical axis.} will produce rings or arcs with an azimuthally dependent photon density. 
Since the morphology and location of the muon ring allow the muon's trajectory and energy to be calculated, it is possible to predict with precision the number of Cherenkov photons that have been collected by the camera. Since the prediction depends only very slightly on the local atmosphere, but goes linearly with mirror reflectivity, transmission of glass coating and the photon detection efficiency of the camera pixels, the detector response to local muons can be calibrated in absolute terms~\cite{vacanti}.

\subsection{Simulations}
\begin{figure}[h]
\centering
\includegraphics[width=0.6\textwidth]{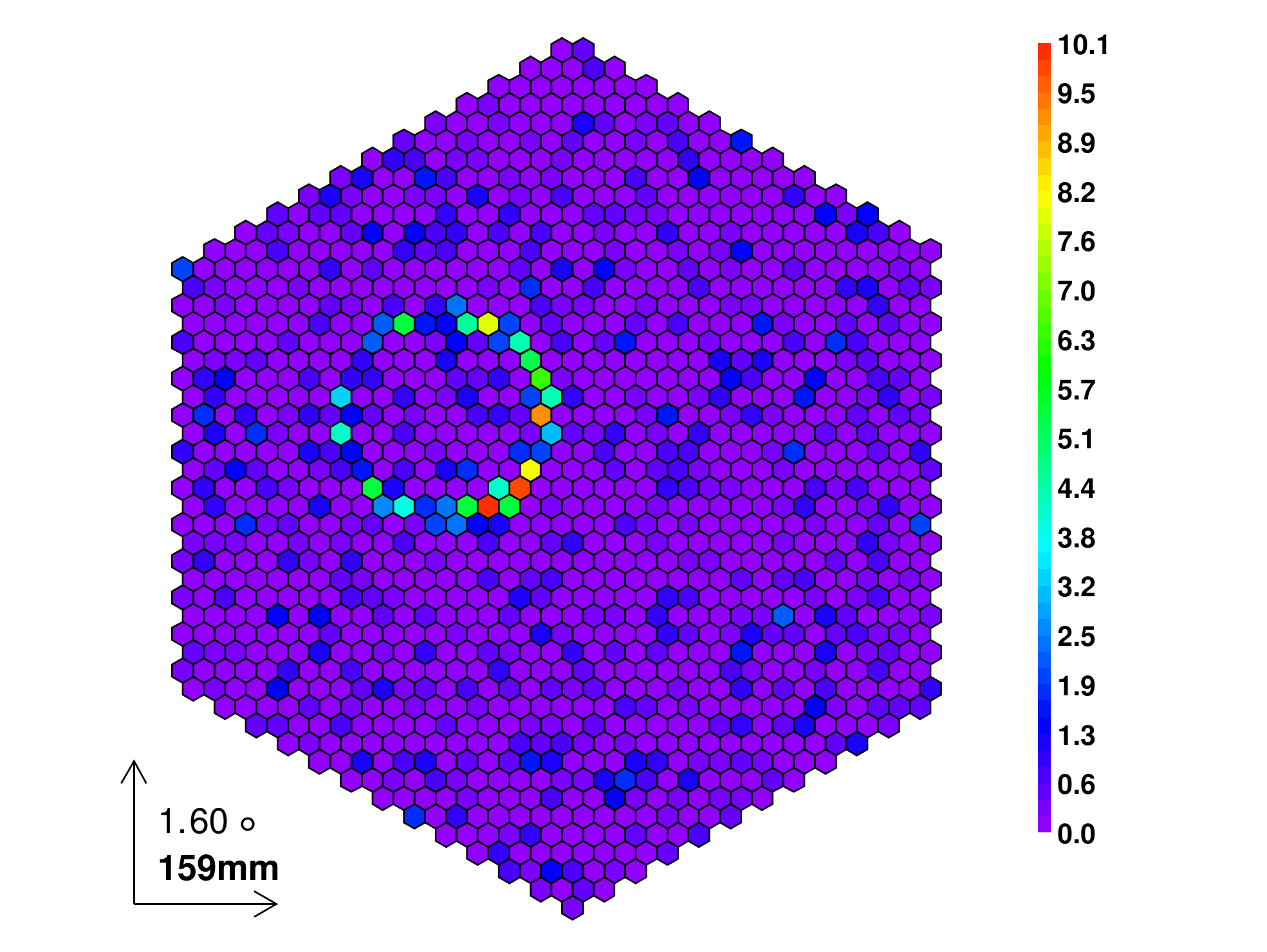}
\caption{Example of a muon image recorded by the SST-1M telescope. The azimuthal non-uniformity is the result of a non-zero impact parameter, while the offset with respect to the centre indicates an off-axis muon direction.}
\label{fig:MuonRingImage}
\end{figure}

The results reported in this contribution have been obtained from simulation studies using standard CTA software, namely CORSIKA \cite{CORSIKA} and the {\tt sim\_telarray} package \cite{SimTel}. The latest telescope and camera parameters (e.g. mirror reflectivity, light collectors, photon detection efficiency of the silicon photo-multiplier sensors forming the SST-1M camera, hexagonal geometry of the camera, etc.) have been included in the simulation of the instrument. The baseline trigger is a digital sum trigger with sampling rate 250~MHz.
Muons  have been injected at a starting altitude of $\sim 1$ km above the observation level\footnote{The reference site for all simulations is \textit{Aar}, the proposed site for the Southern array in Namibia, located at $h_\mathrm{tel} = 1640$~m a.s.l.}, randomly distributed in a cone of $\sim4.5^\circ$ (corresponding to half of FoV of the telescope) around the vertical axis to cover the entire FoV. The telescope has been simulated pointing always to the zenith, and the impact parameter has been randomly distributed from 0 m to $\sim2$ m (radius of the mirror). The muon spectrum has been simulated as a power law with a spectral index of $-2$ in an energy range between 6 GeV and 10 TeV.
An example of a full ring image as it would appear on the SST-1M is shown in Figure~\ref{fig:MuonRingImage}.

\subsection{Analysis of the rings}

The analysis of the muon images requires the calculation of the muon image parameters: the ring radius, the ring width (\textit{ArcWidth}), the opening angle of the ring fraction (\textit{ArcPhi}) and its total light content, measured in photo-electrons (\textit{MuonSize}). 

The analysis has been carried out using MARS, the standard software for the MAGIC telescopes data analysis \cite{MARS}.

\begin{figure}[h]
\centering
\includegraphics[width=0.6\textwidth]{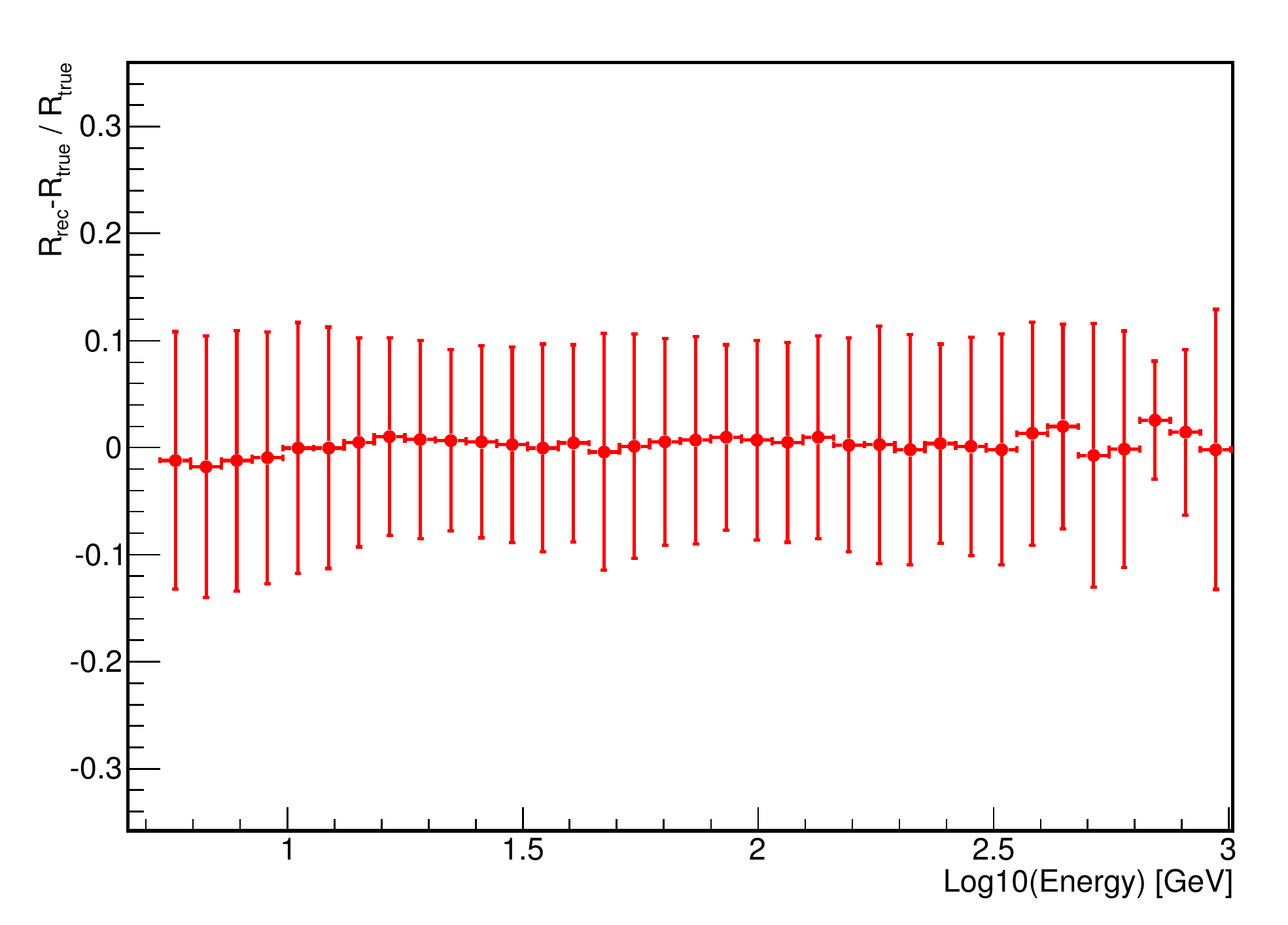}
\caption{Resolution of the muon ring radius as a function of muon energy. The bias on the radius reconstruction is less than $\sim$ 1\%.}
\label{fig:Radius}
\end{figure}

The search for muon rings starts after image cleaning. For this study an absolute image cleaning has been applied with 5 photo-electrons for core and 2 photo-electrons for boundary pixels \cite{MAGICCleaning}. 
After cleaning, every image is fitted by a circle starting at the centre of the Hillas ellipse (Hillas parameters are calculated in a preceding step). From that starting point, the algorithm calculates the distance of the putative image centre to every pixel, and from the distance values the mean (weighted by the pixel content) with its deviation (rms) are calculated. The algorithm minimises the deviation by changing the coordinates of the assumed centre.
Figure~\ref{fig:Radius} shows the resolution of the muon ring radius as a function of the energy, defined as $(R_{rec} - R_{true}) / R_{true}$. The bias on the ring radius reconstruction is less than $\sim$ 1\% for the overall energy range considered and it can be easily subtracted.  

\begin{figure}[h]
\centering
\includegraphics[width=0.6\textwidth]{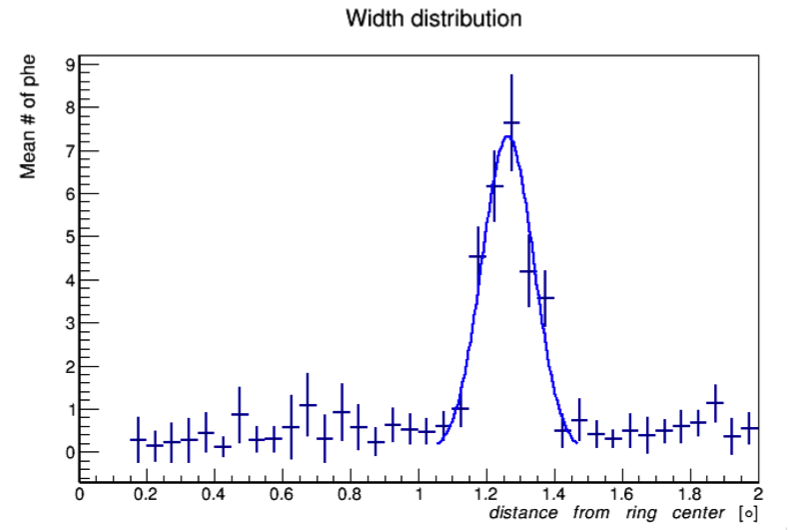}
\caption{Radial intensity distribution with a gaussian fit for the calculation of the \textit{ArcWidth}. The value of the \textit{ArcWidth} is related to (but not exactly) the optical PSF.}
\label{fig:ArcWidth}
\end{figure}

In a second step, the image before cleaning is used to calculate the radial and the azimuthal intensity distribution of the image. For the azimuthal intensity distribution, all pixels inside a certain margin around the radius are used. The following parameters, useful for the analysis can be defined: 
\begin{itemize} 
\item \textit{ArcWidth} is calculated as the sigma value of a Gaussian fit to the signal region in the radial intensity distribution;
\item \textit{ArcPhi} is calculated as the sum of connected bins, which lie above a certain pixel amplitude threshold;  
\item \textit{MuonSize} is calculated as the sum of the contents of all pixels along the ring. 
\end{itemize}
Due to the intrinsic broadening of the ring width, the absolute value of the \textit{ArcWidth} is not equal to the sigma value for the optical PSF, but it is related to that. Therefore, to get the size of the PSF it is necessary to compare simulated muons with different PSFs with muons extracted from observational data. Figure~\ref{fig:ArcWidth} shows an example of the radial intensity distribution of the image with a Gaussian fit to the ring region. 

\section{Study of the optical degradation}

\begin{figure}[h]
\centering
\includegraphics[width=0.48\textwidth]{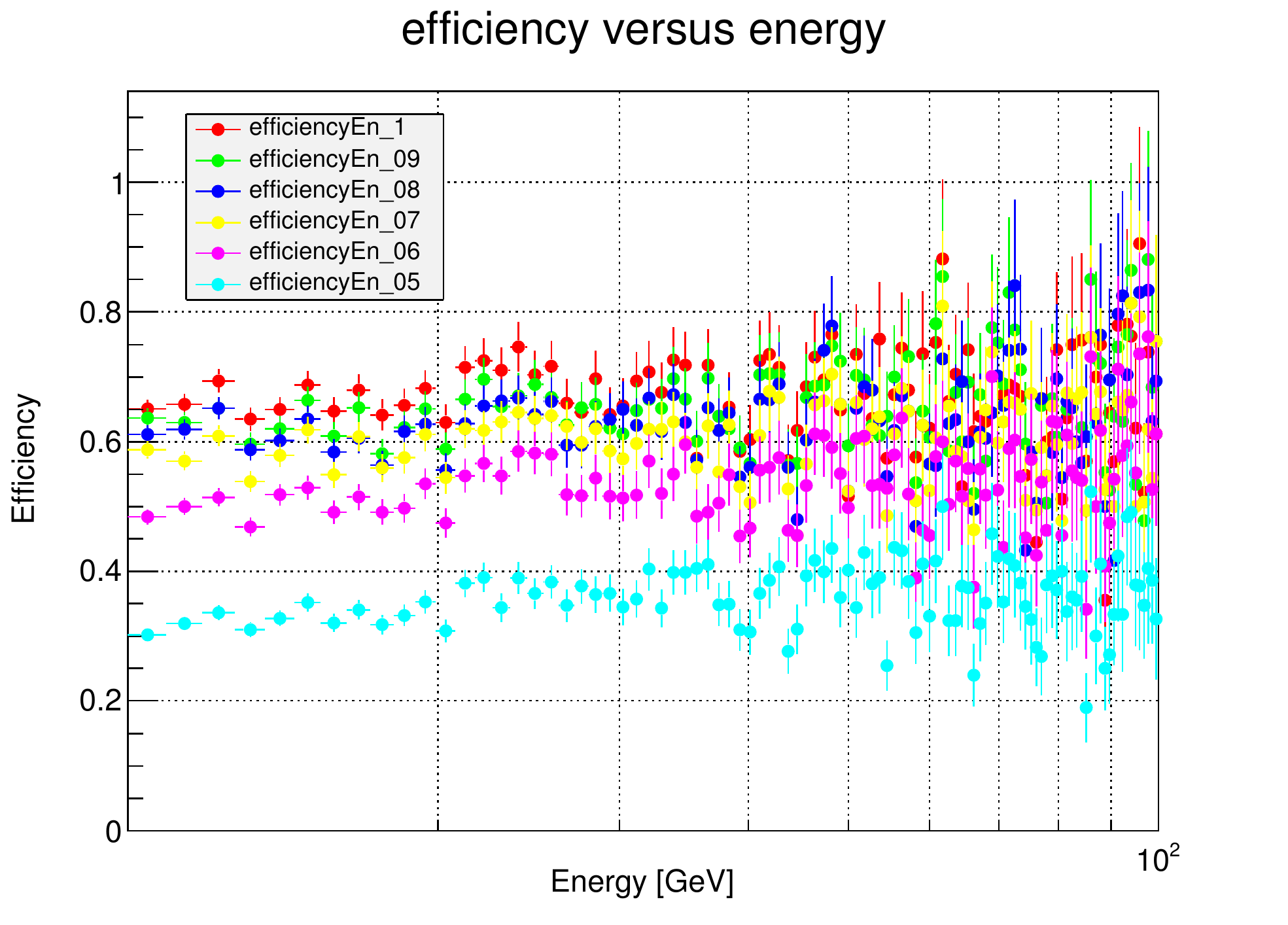}
\includegraphics[width=0.48\textwidth]{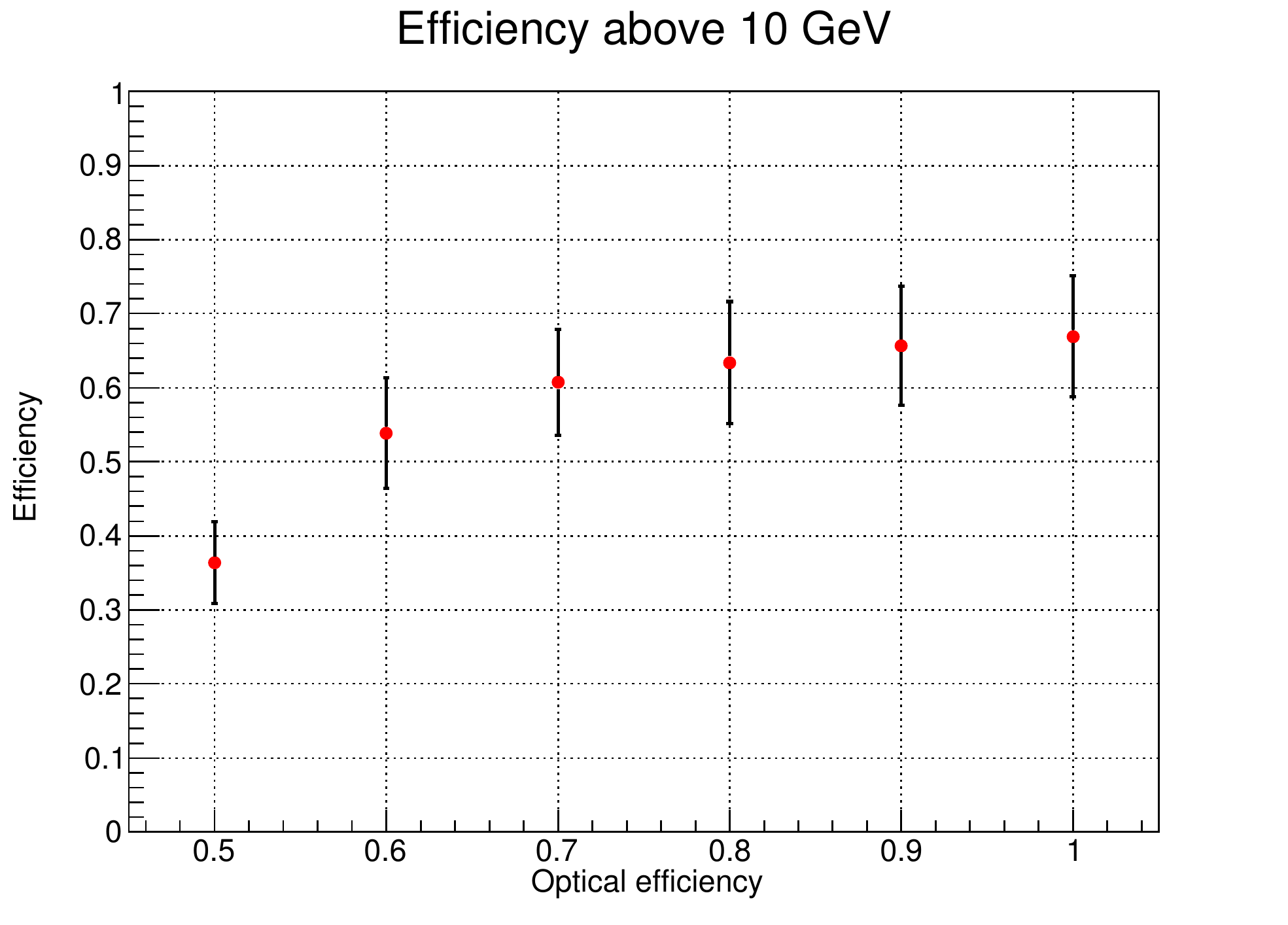}
\caption{Left plot: Muon efficiency for SST-1M obtained with a trigger threshold of 100 ADC counts as a function of energy for different optical efficiencies (100\% in red, 90\% in green, 80\% in blue, 70\% in yellow, 60\% in magenta, and 50\% in cyan). Right plot: Muon efficiency for SST-1M obtained with a trigger threshold of 100 ADC counts and as a function of the optical efficiency for muons with an energy above 10 GeV.}
\label{fig:Efficiency}
\end{figure}

The analysis presented in the previous section has been developed to study the capability of the SST-1M telescope to trigger on muon events and its ability to calibrate the optical throughput, possibly without trigger bias. The effect of the summed pixel amplitude threshold and selection cuts have been investigated with the purpose of obtaining a muon efficiency curve independent from the optical degradation of the telescope. Simulations have been carried out using different scales of mirror reflectivity degradation.  
In order to select good muon events\footnote{Cuts are based on muon simulations only. No study about the cosmic ray rejection has been performed at this stage.} the following quality cuts have been applied based on the previously described analysis: 
\begin{itemize}
\item the ring fitting the muon image is fully contained in the camera;
\item the reconstructed radius is between 0.5$^\circ$ and 1.5$^\circ$;
\item the projected ring width along the ring radius has a gaussian shape (reduced chi square of the \textit{ArcWidth} fit $<$ 2);
\end{itemize}
Figure~\ref{fig:Efficiency} (left) shows the muon efficiency (calculated as the ratio between the observed and simulated muons) curves obtained with a trigger threshold of 100 ADC counts as a function of the muon energy above 10 GeV. The effect of the muon energy on the muon efficiency is rather negligible for all the considered optical efficiencies. 
The right side of Figure~\ref{fig:Efficiency} shows the muon efficiency for energies above 10 GeV as a function of the optical efficiency. The muon efficiency has a constant value of around 60\% in the region of interest (between 60\% to 100\% of optical efficiency). The trigger rate seems to be rather insensitive to the optical efficiency of the system when a lower trigger threshold of 100 ADC counts is adopted\footnote{the general \emph{Prod2} simulation for CTA has been done using a trigger threshold of 127 ADC counts.}.

\section{Conclusions}
Muon rings have been used as a method to calibrate the total optical throughput of practically all previous Imaging Atmospheric Cherenkov Telescopes \cite{HESS, VERITAS, MAGIC}. In this contribution we have presented simulations and analysis of the ring images carried out in the framework of the SST-1M project. This work is done in close collaboration with the other SST teams of the CTA Consortium \cite{GCT, ASTRI}. Preliminary results show that muon images are adequately selected despite a degradation of the optical efficiency of the instrument. Building on this encouraging result the SST-1M collaboration continues to work towards further improving the trigger efficiency for muon rings. In this direction, additional dedicated muon triggers are currently under study.

\section*{Acknowledgements}
We gratefully acknowledge support from the agencies and organizations 
listed under Funding Agencies at this website: http://www.cta-observatory.org/. 
Elisa Prandini gratefully acknowledges the financial support of the Marie Heim-Vogtlin grant of the Swiss National Science Foundation.

\end{document}